# The convergence of chaotic integrals

Oliver Bauer [a,b,c] and Ronnie Mainieri [a]

(a) *Theoretical Division, MS B213, Los Alamos, NM 87545*
(b) *Center for Nonlinear Studies, MS B258, Los Alamos, NM 87545*
(c) *Fachbereich Physik der Universität Regensburg, Institut II 93040 Regensburg, FRG*

## Abstract

We review the convergence of chaotic integrals computed by Monte Carlo simulation, the trace method, dynamical zeta function, and Fredholm determinant on a simple one-dimensional example: the parabola repeller. There is a dramatic difference in convergence between these approaches. The convergence of the Monte Carlo method follows an inverse power law, whereas the trace method and dynamical zeta function converge exponentially, and the Fredholm determinant converges faster than any exponential.

# Introduction

In chaotic systems, the exponential divergence of nearby trajectories makes it pointless to observe individual trajectories for a long time. Instead, collections of orbits are observed and an average value is computed. These averages involve integrating a quantity along a trajectory of a chaotic system — a chaotic integral — and can be computed by different methods. We will discuss different methods to evaluate chaotic integrals and consider how fast, or slow, they converge.

A simple example of a chaotic integral is the computation of the mean displacement squared. If we have a chaotic system where the position of a particle is given by $x$, we can ask what is the average value $\langle x^2 \rangle$, the mean displacement squared. This quantity is given by

$$\langle x^2 \rangle = \frac{1}{T} \int_0^T d\tau \, x^2(\tau) \,, \tag{1}$$

where the value of $T$ must be as large as possible. As written, the value of $\langle x^2 \rangle$ depends on the choice of the initial condition, but for a chaotic system almost every initial value will yield the same result.

The are several ways to evaluate a chaotic integral. The most straightforward method is by Monte Carlo simulation. For a chaotic system this corresponds to choosing a random initial condition, evaluating the integral for some fixed but large $T$, and averaging many trials. The great advantage of the Monte Carlo method is that it assumes very little about the system. The only assumption needed is an ergodicity assumption. Its disadvantage is that it does not converge very fast. Typically, if we have $m$ trails, the convergence goes as $1/\sqrt{m}$. The other methods we will consider require knowing more about the system. They are: the trace method and cycle expansions in the form of zeta functions and Fredholm determinants.

Cycle expansions [1] are a powerful method for computing the asymptotic properties of chaotic systems. They are based on the careful ordering of periodic orbits, using shorter orbits to estimate the effects of longer orbits. Cycle expansions cannot be applied blindly to any system. As input one must know the topology of the periodic orbits, which is the same as the symbolic dynamics for them. This additional input about the system is re-payed in the form of faster convergence of cycle expansions.

Rather than explaining the methods in complete generality we will consider a simple example: that of determining the escape rate of a one-dimensional map. The escape rate is the time it takes a point to leave a certain region of configuration space. It was introduced as a dynamical average by Kadanoff and Tang [2]. We will evaluate the escape rate by the all the methods mentioned: Monte Carlo, trace formula, zeta function, and Fredholm determinant. Although all our simulations are for the escape rate, we expect the convergence results to apply to all classical dynamical averages, such as $\langle x^2 \rangle$, the $f(\alpha)$ spectrum of dimensions, and generalized fractal dimensions. First we will explain how the escape rate is defined and how it can be evaluated by each of the different methods. Then we will compare their convergence rates and conclude that there is a dramatic difference in the convergence rates between the methods.

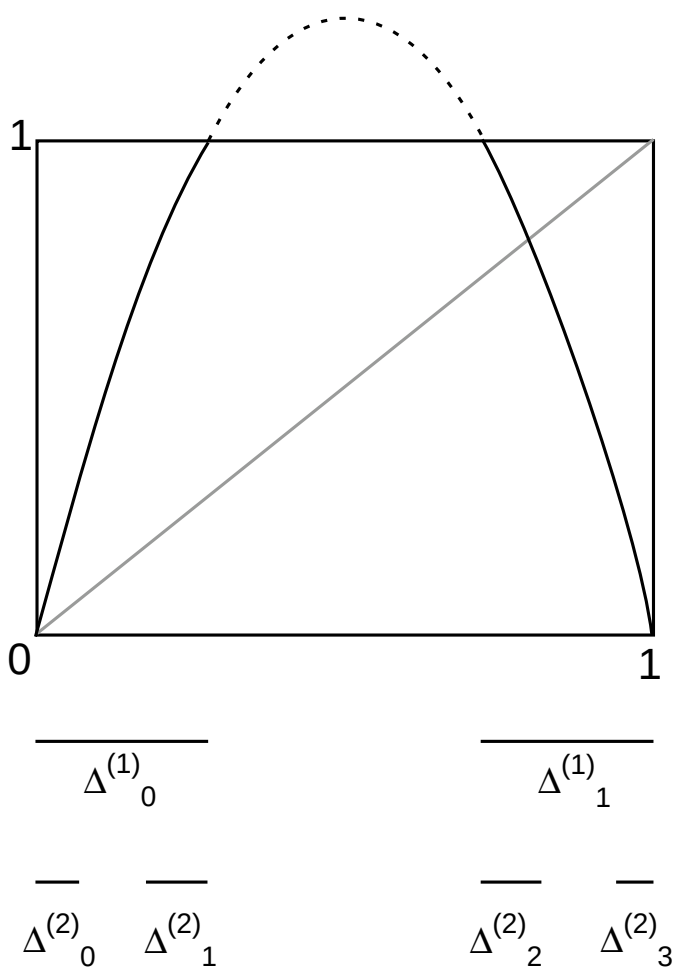

*Figure 1: As the points in the interval* $[0, 1]$ *are iterated by the map , most points escape through the gap in the center. After one iteration only the points in segments* $\Delta_0^{(1)}$ *and* $\Delta_1^{(1)}$ *remain. After two iterations only the points in the segments* $\Delta_k^{(2)}$ *remain.*

## Computing the escape rate

The escape time of a system is a quantity that could be measured in a laboratory. It is related to scattering — it determines the time a particle spends inside a scattering region. Suppose we had a region containing a limited range potential and we scatter a particle off that potential. After a long time the particle will have left the potential region and will once again be in the potential-free region. If, after that long time, an observer compares the radial position of the particle with that of a free particle, the observer will notice that the scattered particle has been delayed. How much it is delayed depends on how many times the particle bounced among the hills of the potential.

The escape rate is the average escape time for a group of particles. We will only consider the discrete time version of the escape rate. To define it in one dimension, consider the parabola map $ax(1-x)$, shown in figure 1. When $a > 4$ not all points of the unit interval get mapped back. Points that get mapped into the gap region (between $(1 \pm \sqrt{1/a - 4/a^2})/2$) never return to the interval, going to $-\infty$ under iteration. When this happens we say that the points have escaped the unit interval. It is the equivalent of leaving the potential region in the earlier example.

Let us now try to determine how many points remain after each iteration. If no iterations have taken place we have the whole unit interval $\Delta_0^{(0)}$. After one iteration all the points in the gap escape the unit interval and we have only two segments with points: $\Delta_0^{(1)}$ and $\Delta_1^{(1)}$. At the next iteration we will have four segment of points remaining; these are the points that did not get mapped into the gap in iteration one. By this construction we see that the segments are determined from the backwards iterations of the gap. The total number of points remaining after $n$ iterations is

$$\Gamma_n = \sum_{k<2^n} |\Delta_k^{(n)}| \ , \tag{2}$$

where we used $|\cdot|$ to indicate the length of the segment. If the map had a constant slope, of say $\Lambda$, each of these segments would be contracting at a rate $\Lambda$, and the typical size of segment $\Delta^{(n)}$ would be $\Lambda^{-n}$. In this case the sum of the segments would be $(2/\Lambda)^n$, as there are $2^n$ segments after $n$ iterations. A map with a gap has $\Lambda > 2$, so the sum would decay exponentially with $n$. We can then define the escape rate $\gamma$ as

$$\gamma = -\lim_{n\to\infty} \frac{\ln \Gamma_n}{n} \ , \tag{3}$$

which gives the rate of escape of points through the gap. If we call $C_n$ the union of all remaining segments after $n$ iterations, then to compute $\Gamma_n$ we can also evaluate the integral of the indicator function of $C_n$

$$\Gamma_n = \int_{[0,1]} dx \, [x \in C_n] = \int_{[0,1]} dx dy \, \delta(y - f^n(x)) \ . \tag{4}$$

We will now review several methods that are used in the chaos literature to compute averages of chaotic systems, such as the escape rate in equation (3). All methods have as their starting point the definition of $\Gamma_n$, equation (4) and use the map $f : x \mapsto ax(1-x)$ with $a = 4.5$. (By choosing $\sqrt{a^2 - 4a} > 1$, which is satisfied if $a > 4.2361$, the derivative of the map is bounded away from 1.) The methods we will review are: the Monte Carlo method, the trace method, zeta functions, and Fredholm determinants.

The simplest method is to evaluate the integral directly by the Monte Carlo method. One simultaneously evaluates the integral (4) for different values of $n$ and then uses the resulting integrals to extrapolate the limit (3) defining the escape rate. To do so, we uniformly distribute $10^{10}$ points on the unit interval. Each point is iterated until it escapes, and the number of iterations recorded. This produces an histogram of escape times, as shown in figure 2. The slope of the histogram is the escape rate. Although in the logarithmic plot in figure 2 the histogram appears as a straight line, it is not. Determining which straight line best fits the data amounts to extrapolating the limit in equation (3). As each point is added to the histogram, the escape rate is computed and plotted, as shown in figure 2.

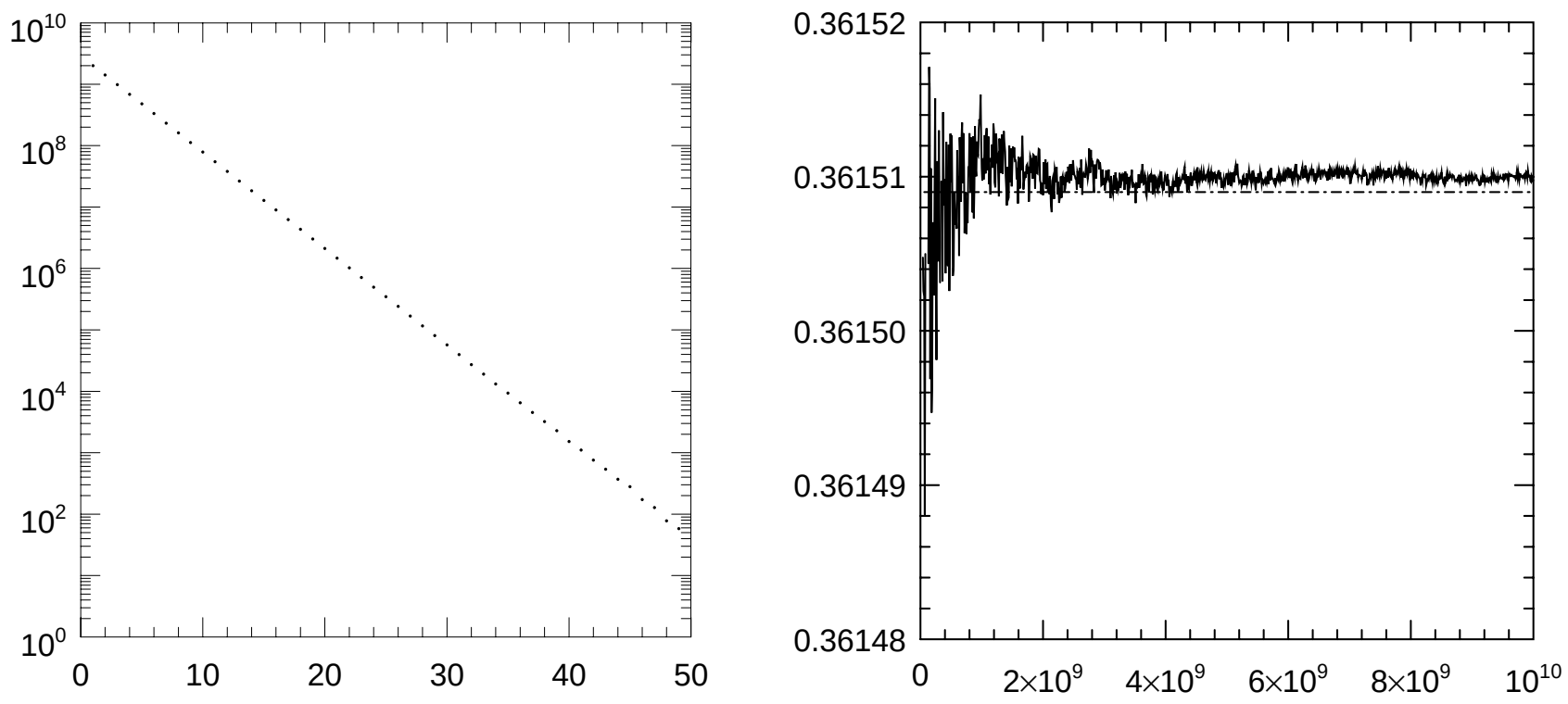

*Figure 2: Convergence of the escape rate when computed with the Monte Carlo method. Despite $10^{10}$ having been used, the escape rate is only determined to 4 digits. The histogram of the number of iterations before escaping is given on the left. After a few points are added to the histogram, one has to accurately determine its slope to obtain the escape rate, which is plotted on the right.*

The convergence of the Monte Carlo method depends on the number $T$ of points iterated and goes as $1/\sqrt{T}$. The escape rate is computed by estimating the slope of the histogram generated from random samples, that is, the slopes themselves are random samples. Their rate of convergence is dictated by the the law of large numbers, which says that the average should converge as $1/\sqrt{T}$. This is very slow and it is difficult to determine if the results have converged or not, as $1/\sqrt{T}$ is a scale invariant function. If we plot a fraction of the computed data on a linear plot the data appear to have converged. As a check one could double the length of the run and verify that the final value has not changed much. This then leads one (as we were mistakenly lead) to conclude that the result obtained had converged.

The reason the Monte Carlo method converges so slowly is that it assumes so little about the system. For the (strong) law of large numbers to apply to $\int f \, dx$ we only need the existence of $\int f^2 \, dx$ (finite variance). In our case both integrals exist. If we use more information about the system, we should expect better convergence.

In the trace method, we notice that $\Gamma_n$ can be written as the trace of an operator. This operator has a largest, isolated eigenvalue, and the escape rate can be computed from it by the power method. (The power method is the repeated application of the operator to a random initial vector; after many iterations the resulting vector will be parallel to the dominant eigenvector.)

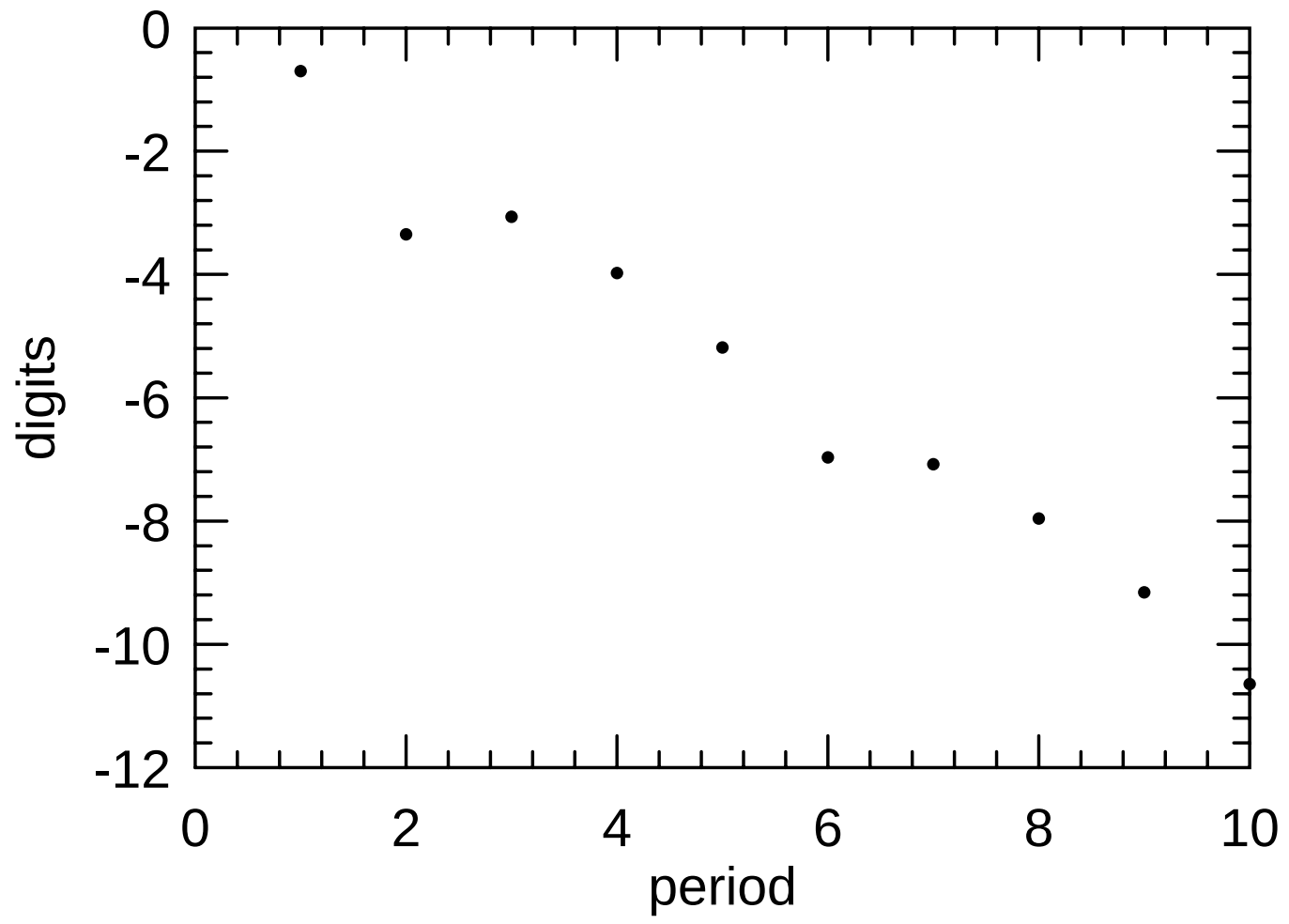

*Figure 3: Convergence of the escape rate $\gamma$ when computed with the trace method: $\gamma = -\frac{1}{n} \ln \operatorname{tr}\mathcal{L}^n$. The escape rate seems to be converging smoothly, but numerical extrapolation methods could not improve the result.*

Introduce the operator $\mathcal{L}$ which acts on functions $g$ of the unit interval

$$\mathcal{L}g(y) = \int_{[0,1]} dx\, \delta(y - f(x))g(x)) \ . \tag{5}$$

The trace of $\mathcal{L}^n$ can be computed easily because of the delta function, and one finds that

$$\operatorname{tr}\mathcal{L}^n = \int dx\, \delta(x - f^{(n)}(x)) \ . \tag{6}$$

Variations on the trace method are used to compute most dynamical averages for chaotic systems, as computing the trace is equivalent to computing a partition function for a fixed size system.

If $\mathcal{L}$ has an isolated eigenvalue, asymptotically the trace depends only on its largest eigenvalue $\lambda_0$, so $\operatorname{tr}\mathcal{L}^n \to \lambda_0^n$. From this we can compute the limit (equation 3) and determine the escape rate to be $\ln \lambda_0$, as explained by Tél [3]. Rather than compute the integral in equation (6) directly, we used the more accessible expression given by equation (12), which will be explained later. This more compact expression simply evaluates the trace of the $\mathcal{L}$ operator to some power $n$. We have done this for different values of $n$ and plotted the resulting values of the escape rate in figure 3. The method does much better than the Monte Carlo method: with 12380 evaluations of the map $f$ we could compute the escape rate to 11 significant decimal digits. This is to be compared to the 4 digits obtained by the Monte Carlo method after $10^{10}$ evaluations of the map.

If we are trying to determine the largest eigenvalue of $\mathcal{L}$, then the trace method is not the most efficient scheme. The trace method depends on com-puting a limit numerically, which is difficult to do accurately. To compute the largest eigenvalue, we could compute the smallest root of the equation

$$\det(1 - z\mathcal{L}) = 0 \ , \tag{7}$$

the characteristic equation for the $\mathcal{L}$ operator. The determinant is interpreted as a Fredholm determinant [4]

$$\det(1 - z\mathcal{L}) \stackrel{\text{def}}{=} \exp\left(\operatorname{tr}\ln(1 - z\mathcal{L})\right) = \exp\left(-\sum_{n>0} \frac{z^n}{n} \operatorname{tr}\mathcal{L}^n\right) \ . \tag{8}$$

We will use the right-most expression to evaluate the Fredholm determinant.

The trace can be computed by carrying out the integral of the delta function. The are only contributions when the argument of the delta function is zero — the fixed points of $f$ composed with itself $n$ times. From equation (6) we have

$$\operatorname{tr}\mathcal{L}^n = \sum_{\substack{x \\ x = f^{(n)}(x)}} \frac{1}{|1 - Df^{(n)}(x)|} \tag{9}$$

where $Df^{(n)}$ is the derivative of the map composed $n$ times:

$$Df^{(n)}(x) = f'(f^{(n-1)}(x)) f'(f^{(n-2)}(x)) \cdots f'(x) \, . \tag{10}$$

A similar expression holds for higher dimensional maps. The expression for the trace is well defined when the hyperbolicity condition holds: the derivative $Df^{(n)}$ must not be 1 and all fixed points must be isolated. For the parabola map we are studying both conditions hold.

To compute the trace of $\mathcal{L}^n$ we have to find all the values of $x$ on the unit interval that are fixed points of $f^{(n)}$. These points return to themselves in $n$ iterations and, therefore, are periodic points of the map $f$. All the fixed points of $f^{(n)}$ can be labeled by the symbolic dynamics of the map [5, 6]. Suppose a point $x$ is a fixed point of $f^{(3)}$, and as it is mapped by $f$, it visits first the left half of the unit interval, then the right, then the left, before returning back to where it started. Its symbolic orbit would be 010, where 0 denotes visiting the left half, and 1 the right half. The symbolic code for each orbit is unique — different symbolic codes correspond to different fixed points. As the symbolic code $\sigma$ determines the starting point $x_\sigma$, we will denote the derivative $Df^{(n)}(x_\sigma)$ by just $\Lambda_\sigma$. The number $\Lambda_\sigma$ is the *stability* of the orbit $\sigma$.

Suppose now that we wanted to determine a list of all the fixed points. We could construct a list of all possible symbolic codes and for each try to determine the fixed point by a numerical method. If we use backward iterations of $f$ rather than forward iterations this is simple to implement.

Two further simplifications in determining all the fixed points are possible. Notice that in formula 9 only the derivative of the map $f$ appears. This means that a fixed point with symbolic code 010 and one with symbolic code 001 will have the same contribution to that sum. That is because $x_{010}$ is in the same periodic orbit as $x_{001}$, as

$$x_{010} \overset{f}{\mapsto} x_{100} \overset{f}{\mapsto} x_{001} \overset{f}{\mapsto} x_{010} \tag{11}$$

and $\Lambda_{010} = Df^{(3)}(x_{010}) = Df(x_{010})Df(x_{100})Df(x_{001}) = Df^{(3)}(x_{001}) = \Lambda_{001}$. Both orbits have the same stabilities, $\Lambda_{010} = \Lambda_{001}$. For the other simplification we notice that if we repeat a periodic orbit k times, the stabilities are multiplied k times. For example, the fixed point $x_{01}$ is also the fixed point $x_{0101}$ and $\Lambda_{0101} = \Lambda_{01}^2$.

With these simplifications we can re-write equation (9), which expresses the trace in terms of periodic orbits. Each orbit class will appear in the sum only once. Rather than including a term with $\Lambda_{001}$, one with $\Lambda_{010}$, and one with $\Lambda_{100}$, only one term will be included and then multiplied by 3, the period $|\sigma|$. One has to be careful with orbits that are repeats of a shorter ones. By the procedure just mentioned the orbit 0101 should contribute four terms, but because it is a repeat of the shorter orbit 01, it only contributes two terms. This is because there are two fixed points $x_{0101}$ and $x_{1010}$ that map into each other. If we restrict our sum in equation (9) to the set $\mathcal{P}$ of *prime orbits*, orbits that are not repeat of shorter ones, we can write

$$\mathrm{tr}\mathcal{L}^n = \sum_{\substack{\sigma \in \mathcal{P} \\ |\sigma| \perp n}} \frac{|\sigma|}{|1 - \Lambda_\sigma^{n/|\sigma|}|} \tag{12}$$

where the sum is over the periodic orbits, and $n = |\sigma| r$ with $r$ being the number of repeats of an orbit. Equation (12) was used for our numerical estimates of the escape rate of the parabola repeller by the trace method. The expression $\mathrm{tr}\mathcal{L}^n$ already converges exponentially, see figure 3. We are now able to evaluate the determinant

$$\begin{aligned} \det(1 - z\mathcal{L}) &= \exp(-\sum_{n=1}^{\infty} \frac{z^n}{n} \mathrm{tr}\mathcal{L}^n) \\ &= \exp(-\sum_{\substack{\sigma \in \mathcal{P} \\ r > 0}} \frac{z^{|\sigma| r}}{r|\sigma|} \frac{|\sigma|}{|1 - \Lambda_\sigma^r|}) \end{aligned} \tag{13}$$

and using the geometrical series with $1/|\Lambda_\sigma^r| < 1$ yields

$$\begin{aligned} \det(1 - z\mathcal{L}) &= \exp(-\sum_{\substack{\sigma, r \\ k \geq 0}} \frac{z^{|\sigma| r}}{r} \frac{1}{|\Lambda_\sigma^r|} \frac{1}{\Lambda_\sigma^{rk}}) \\ &= \exp(\sum_{p,k} \ln(1 - \frac{z^{|\sigma|}}{|\Lambda_\sigma| \Lambda_\sigma^k})), \end{aligned} \tag{14}$$

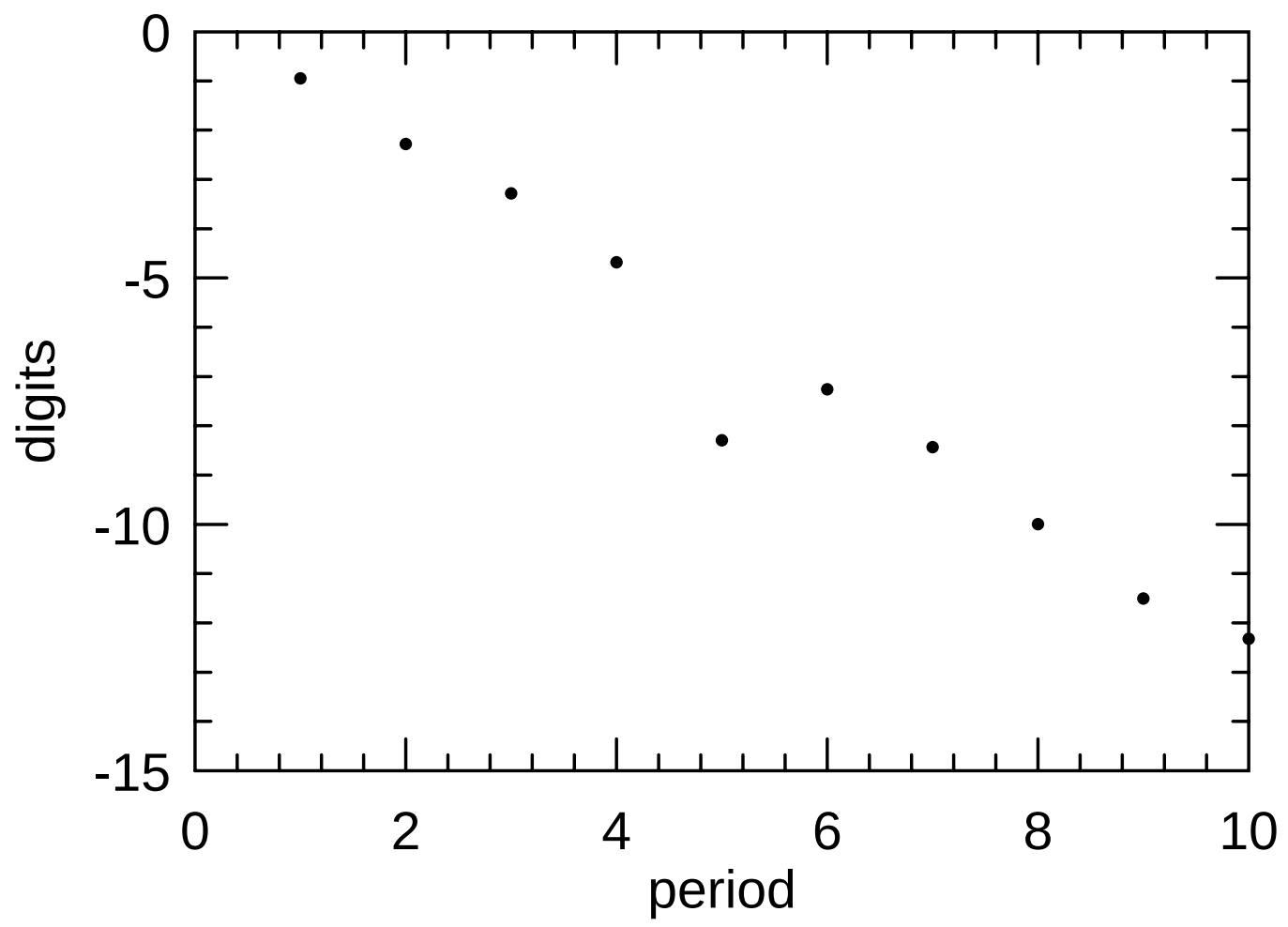

*Figure 4: Convergence of the escape rate when computed with the zeta function method. In this method all orbits of period up to $n$ are used to estimate the escape rate. The stabilities of the orbits are arranged so that the convergence is exponential.*

and finally,

$$\det(1 - z\mathcal{L}) = \prod_{k\geq 0} \prod_{\sigma\in\mathcal{P}} (1 - \frac{z^{|\sigma|}}{|\Lambda_\sigma|\Lambda_\sigma^k}) = \prod_{k\geq 0} \zeta_k^{-1}(z) \ . \tag{15}$$

These derivations can be found in reference [7]

The main theoretical tool in the theory of cycle expansions is the product in equation (15). Its zeros are related to quantities of physical interest [8]. In our case the zero closest to the origin, $z_0$ is related to the escape rate $\gamma$, by $\gamma = \ln z_0$. This zero is a zero of $\zeta_0$ in the product expansion, so in principle one does not need the entire product to obtain the escape rate. As indicated in equation (15), each zeta function is itself a product of many terms.

The results of keeping only the terms for $k = 0$ in the computation of the Fredholm determinant are shown in figure 4. In this case we computed a list of periodic orbits and their stabilities $\Lambda_\sigma$. We then found the smallest root of $\zeta_0$, including all the orbits up to period $n$ when evaluating $\zeta_0$. We let $n$ vary between 1 and 10, displaying in the plot how many digits did not change in the value of the escape rate as orbits of longer period were added. With orbits of period 10 we were able to obtain 12 digits of the escape rate. This was obtained using the same input as the trace method.

The difficulty in using the zeta function $\zeta_0$ is that this function has poles that slow down the convergence of any calculation. These poles may be removed by considering the other zeta functions $\zeta_k$, $k > 0$ [7]. When they are all multi-plied the resulting function is entire [9]. One can see a dramatic change in the convergence of the escape rate when the full product is used. Again in figure 5 we used the same list of orbits and their stabilities as in the computation of the zeta function $\zeta_0$. But rather than using the product form, as in equation (15), we used the exponential form of equation (14); its use entails a smaller number of computer operations. Because of the cancelations involved in evaluating the exponential form, we carried out all our calculations using 70 digit precision arithmetic.

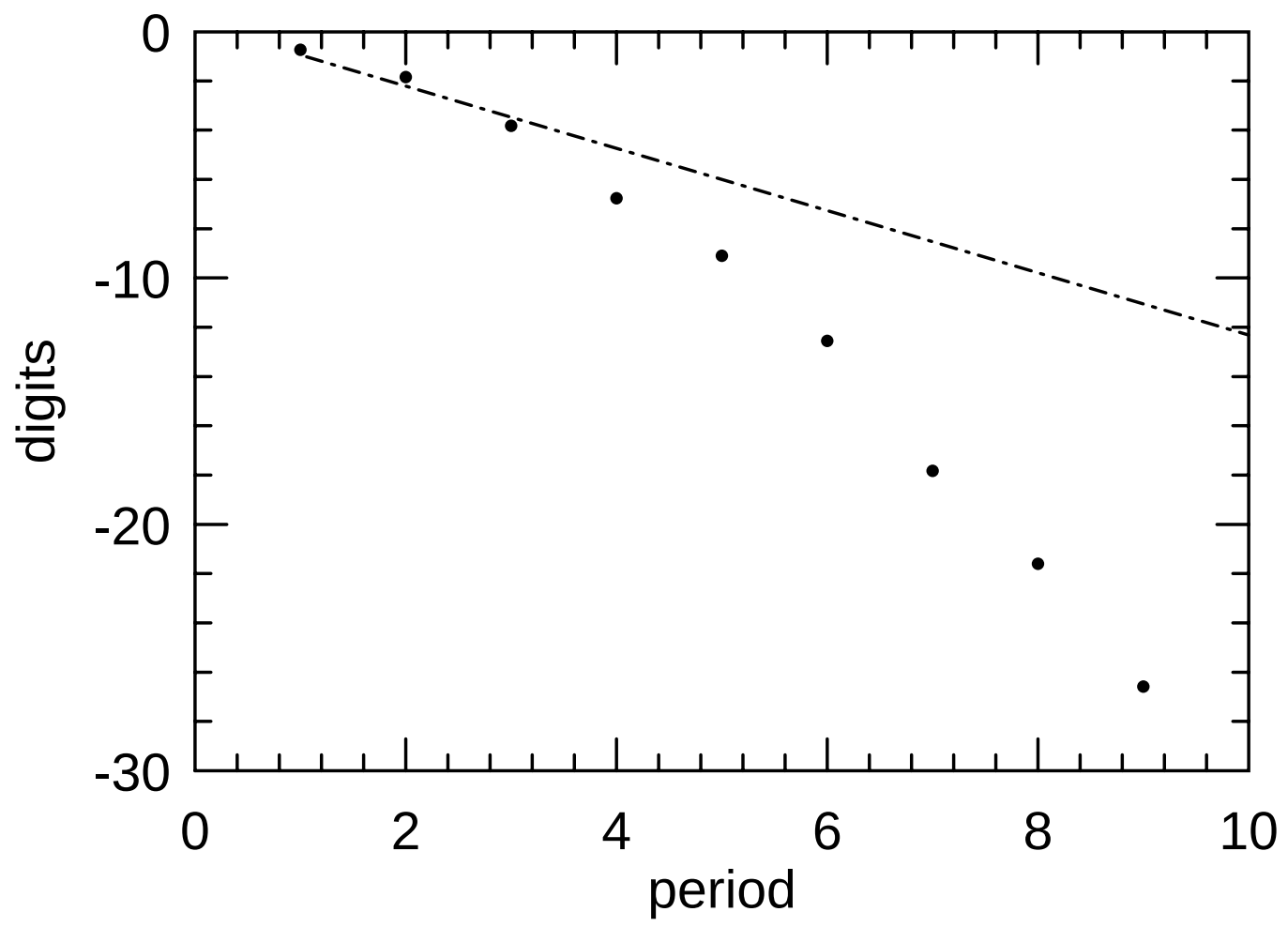

*Figure 5: Convergence of the escape rate when computed with the Fredholm determinant. All orbits of period up to $n$ are used to estimate the escape rate. The Fredholm determinant eliminates the singularities that may exist in the zeta function expansion, and the convergence is faster than any exponential. The exponential convergence rate of the zeta function is indicated as a dashed line.*

# Conclusions

We computed a chaotic integral — the escape rate of a map $f$ — by Monte Carlo simulation, the trace method, with a zeta function, and with a Fredholm determinant. The results are summarized in table 1. The Monte Carlo method evaluated the map $10^{10}$ times and obtained the escape rate to 4 decimal digits. The other methods all use the same input, evaluating the map 12380 times. With this same input the Fredholm determinant computed three times more digits than the trace method (26 digits and 9 digits using orbits of period nine).

Determining the convergence rate of chaotic integrals is a difficult subject, as the rate and type of convergence depends on the type of dynamics and on the observable being averaged. In the Monte Carlo method it depends on the type of sampling being used. In the context of the trace method it was studied numerically by Stoop and Parisi [10] and in terms of the $\mathcal{L}$ operator by Keller [11]. Cvitanović and collaborators have given a series of estimates for the rate of convergence of cycle expansions. Artuso, Aurell, and Cvitanović [7, 12] show that $\zeta_0$ and the Fredholm expansion should converge at least exponentially fast, and later on Cvitanović [13] refined the estimate for the Fredholm determinant and showed that it converges as $\exp(-n^{(d+1)/d})$ for a $d$-dimensional map. A different example, from statistical mechanics, comparing the Monte Carlo method, series expansions, and cycle expansions was given by Mainieri [14].

| Method | Iterations | Escape rate $\gamma$ |
|---|---|---|
| Monte Carlo (2 dimensions) | $10^{10}$ | 0.36 |
| Monte Carlo | $10^{10}$ | 0.3615 |
| Trace | $10^4$ | 0.3615096698 |
| Zeta function | $10^4$ | 0.361509669842 |
| Fredholm determinant | $10^4$ | 0.36150966984203012532793331 |

*Table 1: Numerical results of the various ways of computing the escape rate. The last digit quoted was considered significant. The result for the Monte Carlo method in two dimensions has been divided by two. The convergence rate for different methods are qualitatively different.*

Why do we want to compute the escape rate so accurately? There are no experiments that could measure an escape rate to more than a few digits. The reason we compute it so precisely is to test our understanding of chaotic systems. To compute a quantity precisely one must understand a system well. It is a test. The Monte Carlo method used no information about the system except that it was ergodic; it performed poorly. The Fredholm determinant assumed much more; it performed better. Precision is also necessary when we have to compute higher dimensional chaotic integrals.

The deterioration of accuracy in higher dimensional systems is well illustrated in the Monte Carlo simulation. We re-did the escape rate calculation for a two-dimensional map. As a simple example, we took the map $H : (x, y) \mapsto (f(x), f(y))$, where $f$ is the parabola map we used before (with $a = 4.5$). In figure 6 we have plotted the escape rate as a function of the number of points in the histogram of escape times. The convergence is much slower than the one dimensional case, with a little less than two digits being correct. A simple computation with a Fredholm determinant shows that the escape rate should be $\gamma^{2D} = 0.7230187$, a value that is not even in the plot. Just as before, one has to estimate the escape rate from a histogram, and although there are many points, it is still not in the asymptotic regime. It is easy to see how one can be lead astray by a simulation where one has no way of estimating convergence. While there is little loss of precision in going to higher dimensions in the Fredholm calculation, we expect the Monte Carlo result to worsen with increasing number of dimensions.

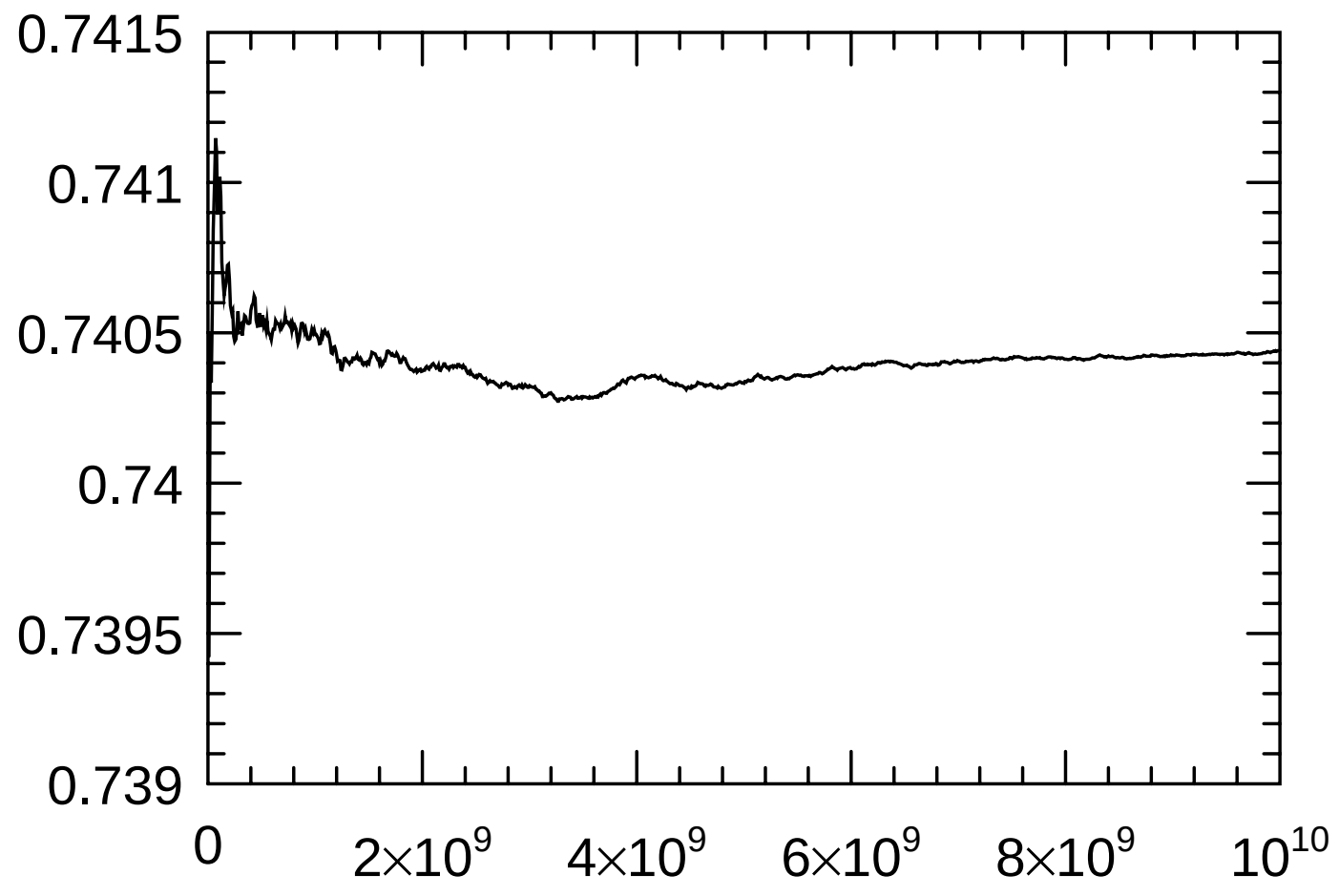

*Figure 6: The escape rate for a two dimensional map. The convergence is slower the higher the dimension of the map. In this case the correct answer is* 0.723*, and not around* 0.7404 *as it appears in the plot.*

We would like to acknowledge the financial support of the Department of Energy.

## References

[1] Predrag Cvitanović. Invariant measurements of strange sets in terms of cycles. *Physical Review Letters*, 61:2729–2732, 1988.

[2] Leo P. Kadanoff and Chao Tang. Escape from strange repellers. *Proceeding of the National Academy of Science, USA*, 81:1276 – 1279, 1984.

[3] T. Tel. Escape rate from strange sets as an eigenvalue. *Physical Review A*, 36:1502–1505, 1987.

[4] Alexandre Grothendieck. *Produits tensoriels topologiques et éspaces nucléaires*, volume 16 of *Memoirs of the American Mathematical Society*. American Mathematical Society, Providence, 1955.

[5] John Milnor and William Thurston. On iterated maps of the interval. In J. C. Alexander, editor, *Dynamical systems : proceedings of the special year held at the University of Maryland, College Park, 1986-87*, volume 1342 of *Lecture notes in mathematics*, pages 465–563. Springer-Verlag, Berlin, 1988.

[6] Kai T. Hansen. *Symbolic dynamics in chaotic systems*. PhD thesis, University of Oslo, 1993.

[7] Roberto Artuso, Erick Aurell, and Predrag Cvitanović. Recyling of strange sets: I. Cycle expansions. *Nonlinearity*, 3:325–359, 1990.

[8] Predrag Cvitanović. Dynamical averaging in terms of periodic orbits. To appear in Physica D., 1995.

[9] Hans. H. Rugh. The correlation spectrum for hyperbolic analytic maps. *Nonlinearity*, 5(6):1237–1263, 1992.

[10] R. Stoop and J. Parisi. Evaluation of probabilistic and dynamic invariants from finite symbolic substrings - comparison between 2 approaches. *Physica D*, 58:325–330, 1992.

[11] G. Keller. On the rate of convergence to equilibrium in one- dimensional systems. *Communications In Mathematical Physics*, 96:181–193, 1984.

[12] E. Aurell. Convergence of dynamic zeta-functions. *Journal of Statistical Physics*, 58:967–995, 1990.

[13] Predrag Cvitanović. Periodic orbits as the skeleton of classical and quantum chaos. *Physica D*, 51:138–151, 1991.

[14] R. Mainieri. Zeta function for the Lyapunov exponent of a product of random matrices. *Physical Review Letters*, 68:1965–1968, March 1992.